\documentclass{aa}
\usepackage{epsfig,epsf,times}

\def\rightleftarrows{\,\raise 0.6 ex\hbox{$\rightarrow$}\kern -1.0 em
 \lower 0.1 ex\hbox{$\leftarrow$}\,}

\newcommand{\mysec}{$.\!\!^{\rm ''}$}

\begin{document}
\thesaurus{08 (02.12.3; 02.18.7; 08.03.4; 09.08.1; 09.10.1)}

\title{Photoevaporation of protostellar disks}
\subtitle{III.\ The appearance of photoevaporating disks around young
intermediate mass stars}

\author{Olaf~Kessel \inst{1,2}, Harold~W.~Yorke \inst{2}
 \and Sabine Richling \inst{2}}

\offprints{O.~Kessel}
\institute{Max-Planck-Institut f\"ur Astronomie, K\"onigstuhl 17,
           D--69117 Heidelberg, Federal Republic of Germany, \\
           Internet: kessel@mpia-hd.mpg.de \and
           Astronomisches Institut der Universit\"at W\"urzburg,
           Am Hubland, D--97074 W\"urzburg, Federal Republic of
           Germany, Internet: yorke@astro.uni-wuerzburg.de, 
           richling@astro.uni-wuerzburg.de} 
\date{Received 10 February 1998 / accepted 26 May 1998}
\maketitle
\markboth{O.~Kessel, et al.: Photoevaporation of protostellar disks III.}
         {O.~Kessel, et al.: Photoevaporation of protostellar disks III.}

\begin{abstract}
We present theoretical continuum emission spectra (SED's), isophotal
maps and line profiles for several models of photoevaporating disks at
different orientations with respect to the observer. The hydrodynamic
evolution of these models has been the topic of the two previous
papers of this series. We discuss in detail the numerical scheme used
for these diagnostic radiation transfer calculations. Our results are
qualitatively compared to observed UCH{\sc ii}'s. Our conclusion is
that the high fraction of ``unresolved'' UCH{\sc ii}'s from the
catalogues of Wood \& Churchwell (\cite{wood}) and Kurtz et
al. (\cite{kurtz}) cannot be explained by disks around massive
stars. In particular, the observed infrared spectra of these objects
indicate dust temperatures which are about one order of magnitude
lower than expected. We suggest that disks around close companions to
OB stars may be necessary to resolve this inconsistency.
Alternatively, strong stellar winds and radiative acceleration could
remove disk material from the immediate vicinity of luminous O stars,
whereas for the lower luminosity sources considered here this will not
occur. We also find that line profiles tracing the evaporated material
originating from the disk are not influenced significantly by the
existence of stellar winds over a wide range of wind velocities (400 -
1000~km\,s$^{-1}$). We compare our results to the bright IRAS source
\object{MWC~349}~A. Many of its properties, especially its spatial appearance in
high-resolution radio maps, can be well explained by a disk
surrounding a {\sc uv} luminous star with a high velocity stellar
wind.


\end{abstract}
\section{Introduction}

A large fraction of early-type main sequence stars are associated with ultracompact H{\sc
ii}-Re\-gions (UCH{\sc ii}s, Wood \& Churchwell \cite{wood}).
These are characterized by large electron densities $N_{\rm e} \ge 10^4$
$\mbox{cm}^{-3}$, sizes $< 0.1$~pc and temperatures $T_e \approx 10\,000$~K.
The overpressure in these regions should lead to expansion and 
dissipation on time scales of a few thousand years. Considering the 
expected lifetimes of massive stars of several million years, however,
the high abundance of observed UCH{\sc ii}s translates into UCH{\sc ii}
mean lifetimes of several $10^5$ years.
This contradiction can be resolved by:
1) the UCH{\sc ii}s could be constrained by high pressure in their
vicinity, or 2) by gravitationally infalling material (Reid et al. \cite{reid}), or 3) there could exist a process which continuously
``feeds'' the UCH{\sc ii}s with matter.
High pressures certainly can be expected in the highly turbulent
molecular cloud cores, which are the birthplaces of young massive stars
(De~Pree et al.~\cite{depree}, Garc\'{\i}a-Segura \&\ Franco~\cite{segura}, Xie et al. \cite{xie}).
Still, it is not clear how turbulence in a cold clumpy medium can contain
the warm ($T \sim 10^4$~K), high density ionized material for extended
periods of time -- many of the technical details of this proposal need
to be worked out.

The photoevaporating disk model proposed by Hollenbach et al.
(\cite{hollenbach93}) and Yorke \&\ Welz (\cite{yowe93}) offers an 
attractive alternative. A circumstellar disk around a luminous OB star 
is continuously photoionized by the central source. The existence of
a powerful stellar wind can modify the quantitative details of
this model, but the basic result remains the same.
Long-lived UCH{\sc ii}s are the necessary
consequence of disks around hydrogen-ionizing sources. In a subsequent
paper by Hollenbach et al. (\cite{hollenbach94}) the quasi-steady state
structure of disks a\-round ionizing sources with winds has been calculated
\mbox{(semi-)} analytically and in Yorke (\cite{yorke95}),
Yorke \&\ Welz (\cite{yowe96}, hereafter Paper I), and Richling
\&\ Yorke (\cite{riyo}, hereafter Paper II) the evolution of such 
circumstellar disks has been followed numerically under a variety of
conditions. 

In Paper I it has been stressed that the phenomenon of disks in the
process of photoionization is not restricted to the (pre\-sum\-ably highly
symmetrical) case of circumstellar disks a\-round OB stars. Disk
formation is a common by-product of the star formation process.
Because OB stars seldom form in isolation, close companions with disks
to a powerful source of ionizing {\sc uv} radiation and a stellar wind 
should be common. Strongly asymmetric UCH{\sc ii}s should result.

Wood \&\ Churchwell~(\cite{wood}) observed 75 UCH{\sc ii}s at
$\lambda=2$~cm and 6~cm with spatial resolution of 0\mysec4 using the
VLA telescope and classified them by their spatial morphological
structure into several types:
\begin{itemize}
\item cometary shaped (20\%),
\item core-halo (16\%),
\item shell type (4\%),
\item irregular or multiply peaked (17\%) and
\item spherical or unresolved (43\%).
\end{itemize}
In order to interpret these observations in light of the photoionized
disk models, further work must be done in refining the hydrodynamical
models for the asymmetric morphological configurations expected when 
a disk is ionized by external sources. Diagnostic radiation transfer
calculations of these numerical models are necessary for a quantitative
comparison.

Goal of the present investigation is to determine spectral characteristics 
and to calculate the expected isophote maps of the {\em symmetrical} 
UCH{\sc ii}s which result from circumstellar disks around OB stars.
We are restricted by the limited number of star/disk configurations 
which have been considered to date. We discuss in detail the physical 
(Sect.~\ref{physmod}) and numerical (Sect.~\ref{nummod}) models
of radiation transfer which we employed. The results for selected
hydrodynamical models from Papers I and II are discussed in
Sect.~\ref{results} and compared to observations of specific sources
in Sect. \ref{comparison}. We summarize our main conclusions in Sect.~\ref{sect:conclusions}.

\section{The physical model} \label{physmod}
In Papers I and II of this series the time dependent photo\-e\-vap\-o\-ration 
of a 1.6~M$_{\odot}$ circumstellar disk around a 8.4~M$_{\odot}$ star 
was calculated under a variety of physical conditions.
The ionizing flux of the central source and its ``hardness'' 
as well as the stellar wind parameters (mass loss rate and terminal velocity)
were varied. States of these models at selected evolutionary times are the basis for our diagnostic radiation
transfer calculations.

\subsection{Continuum transport}

To determine the continuum spectral energy distribution (SED) over a 
frequency range from the radio region up to the optical, we take into 
account three major radiation processes:
thermal free-free radiation (i.e.\ bremsstrahlung of electrons
moving in the potential of protons in the H\,{\sc ii}-region),
thermal dust radiation
and the radiation emitted from the photosphere of an embedded source.

\subsubsection{Free-free radiation}

For this process we adopt the approximation for the emission coefficient
(Spitzer~\cite{spitzer}):
\begin{equation} \label{eq:kapff}
  \epsilon_{\rm ff}=\frac{8}{3} \left( \frac{2 \pi}{3} \right)^{\frac{1}{2}}
  \frac{e^6}{m^2c^3} \left( \frac{m}{kT_{\rm e}} \right)^{\frac{1}{2}} g_{\rm
  ff} N_{\rm e} N_{\rm p} \exp \left(-\frac{h\nu}{kT_{\rm e}} \right) .
\end{equation}
Here, $N_{\rm e}$ and $N_{\rm p}$ are the particle 
densities of electrons and protons. All other symbols have their usual
meanings. We approximate the
Gaunt factor $g_{\rm ff}$ for a non-relativistic plasma by:
\begin{equation}
  g_{\rm ff}=\mbox{max} \left\{ \frac{3^{1/2}}{\pi} \left( \ln
  \frac{\left(2kT_{\rm e}\right)^{3/2}}{\pi e^2 \nu m^{1/2}}-\frac{5
  \gamma}{2}\right) ,1 \right\} ,
\end{equation}
where $\gamma$ ($\approx 0.577$) is Euler's constant.
Assuming the validity of Kirchhoff's Law
$S_{\nu} = \epsilon_\nu / \kappa_\nu = B_{\nu}(T_{\rm e})$,
the absorption coefficient for thermal free-free radiation can be written
($h\nu \ll kT_{\rm e}$):
\begin{equation}
  \kappa_\nu^{\rm ff} = \frac{4 (2\pi)^{1/2}e^6N_{\rm e} N_{\rm p} g_{\rm
  ff}}{(3 m k)^{3/2} c T_{\rm e}^{3/2} \nu^2} .
\end{equation}

\subsubsection{Dust emission}

We adopt the `dirty ice' dust model developed by
Preibisch et al.~(\cite{preib}), which
includes two refractory components: amorphous carbon grains (aC) and silicate
grains as well as volatile ice coatings on the surface of the silicate
grains at temperatures below 125 K (Core Mantle
Particles, CMP's). The icy coatings contain 7\%\ of the available
amorphous carbon and consist of water and ammonium with a volume
ratio of 3:1. At temperatures above 125~K the silicate core and approximately
11 amorphous carbon particles are released into the dusty gas for each CMP.
In Table~\ref{tab:dust} the sublimation temperature $T_{\rm sub}$,
the mean radius $\bar{a}_{\rm d}$ and the number of grains per gram
gas $n_{\rm d}$ are listed for the different species.
\begin{table}[tb]
 \caption[Staubparameter]{Parameters for the grain species used in the
dust model of Preibisch et al.~(\cite{preib}).}
  \begin{flushleft}

 \begin{tabular}{llll}
   \hline\noalign{\smallskip}
   Grain Species & $T_{\rm sub}/[\mbox{K}]$
  & $\log \bar{a}_{\rm d}/[\mbox{cm}]$ & $\log
   n_{\rm d}/[\mbox{g}^{-1}]$ \\
   \noalign{\smallskip}
   \hline\noalign{\smallskip}
   aC & 2000 & $-$6.024 & 14.387 \\
   Silicate & 1500 & $-$5.281 & $-$ \\
   CMP & 125 & $-$5.222 & 12.218 \\ 
  \noalign{\smallskip}
  \hline
 \end{tabular}
 \label{tab:dust}
 \end{flushleft}
\end{table}

\begin{figure}
\begin{center}
\epsfig{file=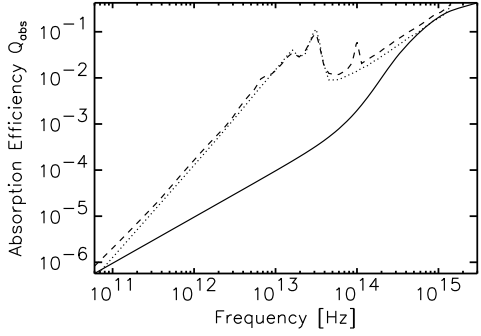,height=5.02cm}
\end{center}
\caption[]{Mean absorption efficiencies for the different dust components.
 Solid line: amorphous carbon, dotted line: silicate, dashed line: CMP's.}
\label{fig:dustparms}
\end{figure}

The absorption coefficient [${\rm cm}^{-1}$] 
for the individual dust components is given by:
\begin{equation}
  \kappa_\nu^{\rm d} = n_{\rm d} \rho \pi \bar{a}_{\rm d}^2
      Q^{\rm abs}_{\rm d,\nu} \; ,
\end{equation}
where the mean absorption efficiency $Q^{\rm abs}_{\rm d,\nu}$ for grain type
``d'' has been determined
using Mie theory for spherical grains of a given size distribution.
Figure~\ref{fig:dustparms} displays the absorption efficiencies for the
different dust components as a function of frequency.
Each dust component's contribution to the source function due to
thermal emission $S_{\nu}^{\rm d}$ is also calculated under the assumption
that $S_{\nu}^{\rm d} = B_{\nu}(T_{\rm d})$.

\subsubsection{Net continuum absorption and emission}

Both emission processes mentioned above occur simultaneous\-ly within the same
volume. Thus the net absorption coefficient and source function are:
\begin{equation}
  \kappa_\nu = \sum_{\rm d} \kappa^{\rm d}_\nu + \kappa^{\rm ff}_\nu
\end{equation}
\begin{equation}
  S_\nu = \frac{1}{\kappa_\nu}
  \left( \sum_{\rm d} \kappa^{\rm d}_\nu B_\nu(T_{\rm d})
  + \kappa^{\rm ff}_\nu B_\nu(T_{\rm e}) \right) .
\end{equation}

\subsection{Forbidden lines}

In order to calculate profiles of the forbidden lines for the elements oxygen
and nitrogen ([O\,{\sc ii}] 3726, [O\,{\sc iii}] 5007 and [N\,{\sc ii}] 6584),
we adopt the following procedure.
First, the equilibrium ionization structure of these elements is 
calculated over the volume of consideration. Next, the
occupation densities of me\-ta\-stable levels $N_{\rm i}$ due to collisional
excitation by electrons is determined. We take into account Doppler shifts
due to bulk gas motions and thermal Doppler broadening to
calculate the profile function $\phi$: 
\begin{equation}
  \phi(\nu) = \frac{1}{\sqrt{\pi} \Delta \nu_{\rm D}} \exp \left[ -\left(
  \frac{\nu - \tilde{\nu}}{\Delta \nu_{\rm D}} \right)^2 \right]  ,
\end{equation}
with the thermal Doppler width:
\begin{equation} \label{eq:doppler}
  \Delta \nu_{\rm D} = \frac{\tilde{\nu}}{c}\sqrt{\frac{2RT_{\rm e}}{\mu}}
    .
\end{equation}
Here $R$ is the gas constant, $\mu$ the atomic weight of the relevant ion,
and $\tilde{\nu} = \nu_{\rm 0}(1+v_{\rm R}/c)$ is the transition
frequency $\nu_{\rm 0}$ Doppler-shifted by the radial velocity
$v_{\rm R}$ of the gas relative to the observer.

The emission coefficient of the transition $k \rightarrow j$, which
enters into the equation of radiative transfer, is then given by:
\begin{equation} \label{eq:lineemis}
  \epsilon_{\rm L}(\nu) = \frac{1}{4 \pi} N_{\rm k} A_{\rm kj} h \nu_{\rm 0}
  \phi(\nu) = \tilde{\epsilon}_{\rm L} \cdot \phi(\nu)  ,
\end{equation}
where $A_{\rm kj}$ is the Einstein coefficient for spontaneous emission.
Note that we have neglected radiative excitation and stimulated emission
in this approximation.

\subsubsection{Ionization equilibrium}

The equations for ionization equilibrium for two neighboring
ionization stages, ${\rm r}$ and ${\rm r}+1$, are:
\begin{eqnarray} \label{eq:ionglg}
  N^{\rm r} \left[ \int_{\nu_{\rm I}}^\infty f_\nu \sigma_\nu^{\rm r} \mbox{d} \nu
  +N_{\rm e} q^{\rm r}+N_{\rm p} \delta^{\rm r} \right] = \nonumber \\
N^{{\rm r}+1} \left[ N_{\rm e} \left( \alpha_{\rm R}^{\rm r}+\alpha_{\rm D}^{\rm
  r} \right) + N_{{\rm H}^{\rm 0}} \delta'^{\rm r} \right]  .
\end{eqnarray}
We solve these equations simultaneously for the $N^{\rm r}$ up to the
ionization stage $r=3$ for both oxygen and nitrogen.
 
\paragraph{Radiative ionization.} The rate of radiative ionization is
calculated from the flux of incident photons $f_\nu$ and the absorption
cross section $\sigma_\nu^r$ integrated over all ionizing frequencies.
We use the radiation field of the central source and neglect scattering
to determine $f_\nu$:
\begin{equation}
  f_\nu = \frac{1}{h\nu} \frac{B_\nu(T_*)
  R_*^2 \pi}{4 \pi R^2} \exp (-\tau)  .
\end{equation}
An analytical expression for the absorption cross section
$\sigma_\nu^r$ is given in Henry~(\cite{henry}).

\paragraph{Collisional ionization.} This ionization process is important in hot
plasmas, where the mean kinetic energy of the electrons is comparable to the
ionization potentials of the ions. N\,{\sc i}, for example, has an ionization
potential 14.5~eV; the corresponding Boltzmann temperature is
$\sim$~170\,000~K. The coefficient for collisional ionization
$q^{\rm r}$ is approximated
by the analytical expression in Shull \&\ van Steenberg~(\cite{shull}).

\paragraph{Radiative recombination.} This is the inverse process to
radiative ionization. For the recombination coefficient
$\alpha_{\rm R}$ we use the formula given in Aldrovandi \&\
Pequinot~(\cite{aldro1}, \cite{aldro2}).

\paragraph{Dielectronic recombination.} 
The probability for recombination is
enhanced when the electron being
captured has a kinetic energy equal to the energy necessary to excite a
second electron in the shell of the capturing ion. The density of excited
levels in the term scheme of the ions grows with energy. Thus, this process
becomes more and more important with increasing temperature. We use two
analytical expressions for $\alpha_{\rm D}$: one for temperatures between
2000~K and 60\,000~K (Nussbaumer \&\ Storey~\cite{nuss}) and one for higher
temperatures (Shull \&\ van Steenberg~\cite{shull}).
\paragraph{Charge exchange.} \label{sect:chargeex} The exchange of electrons during encounters of
atoms and ions, e.g. $N^{++}+H^{\rm 0} \rightarrow N^++H^+$ is also
important. Arnaud \&\ Rothenflug~(\cite{arnaud}) give an expression for the
coefficients $\delta'^{\rm r}$. Special care is necessary in the case of the
reaction $O^+ + H^{\rm 0} \rightleftarrows O^{\rm 0}+H^+$. Due to the similarity
of the ionization energies of hydrogen and oxygen ($\Delta E=0.19$~eV) the
backward reaction is also very effective. At sufficiently high electron
temperatures this leads to the establishment of an ionization ratio
$ N_{O^{\rm 0}}/N_{O^+} \approx (9/8) N_{H^{\rm 0}}/N_{H^+} $,
even in the absence of ionizing radiation. We explicitly include both reactions
in Eq.~(\ref{eq:ionglg}) via the term $\delta^{\rm r}$. An expression
for this coefficient can also be found in Arnaud \&\ Rothenflug~(\cite{arnaud}).

\subsubsection{Collisional excitation of metastable states}

Neglecting the effects of radiative excitation and stimulated emission,
we solve the equations of excitation equilibrium for the population
densities $N_{\rm k}$ (sums over all values ``j'' for which the
conditions under the summation signs are fulfilled):
\begin{eqnarray}
  N_{\rm k} \left[ N_{\rm e} \sum_{E_{\rm j} \neq E_{\rm k}} q_{\rm kj}+
  \sum_{E_{\rm j}<E_{\rm k}} A_{\rm kj}
  \right] & = & \nonumber \\
  N_{\rm e} \sum_{E_{\rm j} \neq E_{\rm k}} N_{\rm j} q_{\rm jk} & + &
  \sum_{E_{\rm j}>E_{\rm k}} N_{\rm j} A_{\rm jk}  ,
\end{eqnarray}
together with the condition $ \sum N_{\rm j} = N_{\rm ges}$.
We use the formulae for the activation and deactivation coefficients given in
e.g.\ Osterbrock~(\cite{oster}):
\begin{equation}
  q_{12}=8.63 \cdot 10^{-6} \, \frac{\Omega_{12}}{\omega_1} \, T_{\rm e}^{-1/2} \exp
  \left(- \frac{\Delta E_{12}}{k T_{\rm e}} \right)
\end{equation}
and
\begin{equation}
  q_{21}=8.63 \cdot 10^{-6} \, \frac{\Omega_{12}}{\omega_2} \, T_{\rm e}^{-1/2}
   ,
\end{equation}
where $\Omega_{12}$ denotes the collision strength for the transition
$1 \rightarrow 2$, $\omega_1$ and $\omega_2$ the statistical weights
of both states involved and $\Delta E_{12}$ the energy difference
between them. For the $\Omega_{12}$ we use the tables given in
Osterbrock~(\cite{oster}).

\subsection{Balmer lines}

Our neglect of line absorption of Balmer photons by hydrogen is justified
as long as the density of Ly$_\alpha$ photons is sufficiently 
low to insure that the hydrogen 2p state is not significantly populated.
This is equivalent to the assumption that Ly$_\alpha$ photons generated
in the nebula by recombination either are quickly destroyed, e.g.\ by dust
absorption or by hydrogen Ly$_\alpha$ absorption followed by 2-photon
emission, or are able to escape sufficiently rapidly, e.g.\ by a random
walk in frequency (Osterbrock~\cite{oster2}).
The emission coefficient of the Balmer lines is given by:
\begin{equation}
  \tilde{\epsilon}_{\rm L}({\rm H}_{\rm i})
  = \frac{1}{4 \pi} \alpha_{{\rm H}_{\rm i}}^{\rm eff}
    \cdot N_{\rm p} N_{\rm e} h \nu_{{\rm H}_{\rm i}}  .
\end{equation}
The effective recombination coefficients $\alpha_{{\rm H}_{\rm i}}^{\rm eff}$
used in this work were adopted from Hummer \&\ Storey~(\cite{hummer}).

\subsection{Radiation from the central star}

As argued in Paper I the resulting {\sc uv} spectrum of a star
accreting material via an accretion disk is very uncertain.
For simplicity we have assumed that the photospheric emission of the central
source (star + transition zone) can be approximated by a black body of given
temperature $T_*$ in the frequency range of interest ($\lambda < 100$~nm).
$T_*$ determines the ``hardness'' of the ionizing photons, thus affecting
both the nebula temperature and the ionization fraction of oxygen and nitrogen.
We use the same values for $T_*$ as in Papers I and II for the hydrodynamic
models.

Nevertheless, the successful spectral classification of the ionizing
star in the UCH{\sc ii} region G29.96-0.02 by Watson \&\ Hanson
(\cite{watson}) gives rise to the hope that more information on
the spectral properties of young, still accreting massive stars
will be available in the future.
 
\section{The Numerical Model} \label{nummod}

\subsection{Structure of the underlying models} \label{intmod}

\begin{figure}
\epsfig{file=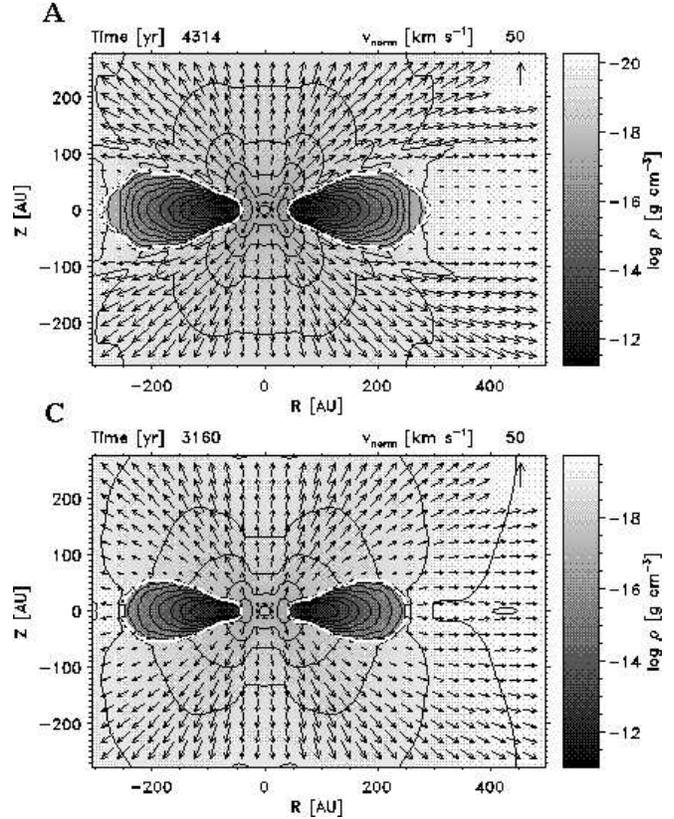,width=8.6cm}
\caption[]{Density, velocity and ionization structure of model A and C.
 Gray scale and black contour lines display the density structure. These
 contour lines vary from $\log\rho=-13.0$ to $\log\rho=-19.5$ in
 increments of $\Delta\log\rho=0.5$. The white contour lines mark the
 position of the ionization front and the arrows show the velocity
 field. The normalization is given at the upper right corner.}
\label{fig:acmodels}
\end{figure}

\begin{figure}
\epsfig{file=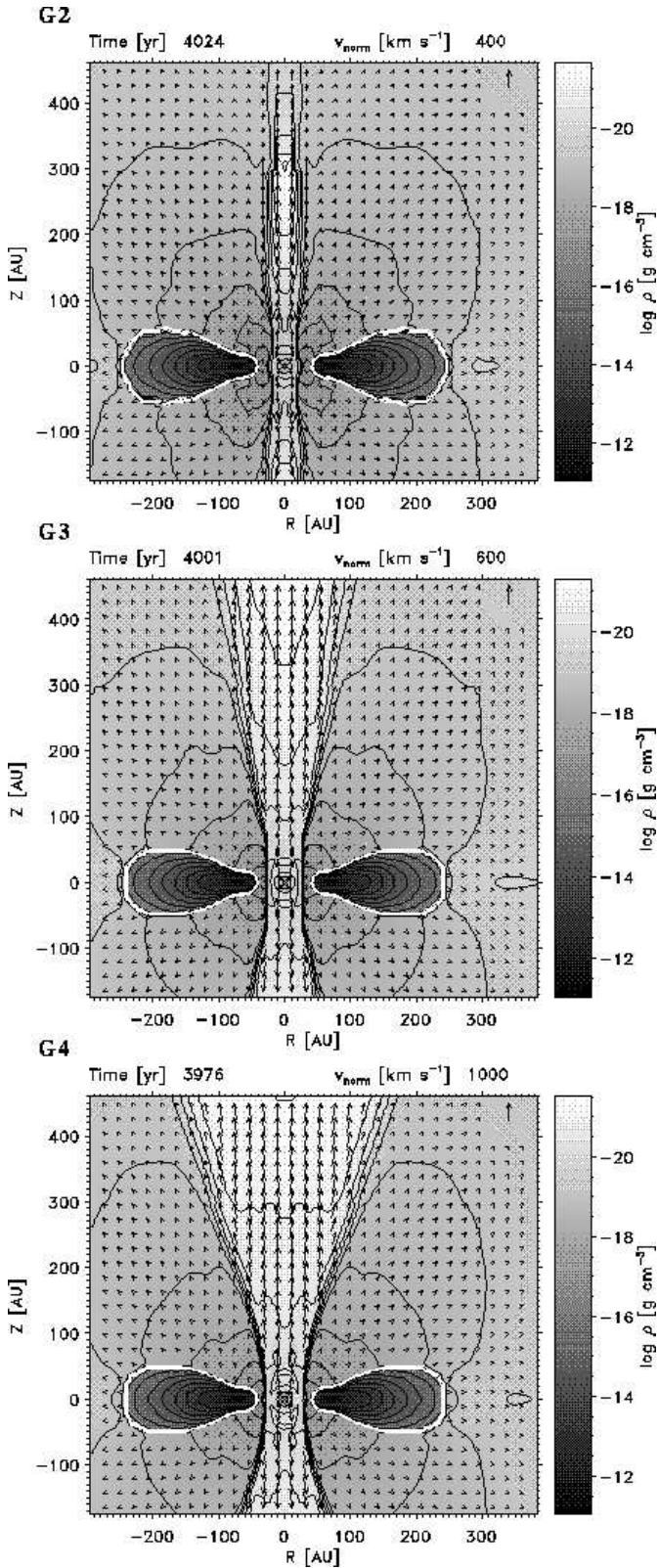,width=8.8cm,clip=1}
\caption[]{Density, velocity and ionization structure of model G2, G3 and
 G4. Symbols and lines have the same meaning as in Fig.~\ref{fig:acmodels}
 except the black density contour lines, which are drawn down to 
 $\log\rho=-21.5$.}
\label{fig:gmodels}
\end{figure}

The underlying numerical models were calculated on five multiply
nested grids, each with 62 x 62 grid cells (see Yorke \&\
Kaisig~\cite{yokai}, Paper I, and Paper II). The spatial resolution
of the finest grid was $\Delta R = \Delta Z \approx 2 \times
10^{13}$~cm (R is the distance to the symmetry axis, Z to the
equatorial plane). Axial symmetry and mirror symmetry with respect to
the equatorial plane were assumed for the models. The simulations were
performed within a volume $(R_{\rm max}^2 + Z_{\rm max}^2)^{1/2} \le
10^{16}$~cm until a quasi-steady state was reached.

For the diagnostic radiation transfer calculations discussed
here we use the final states of five simulations described in Paper
II. Some of the relevant parameters of these simulations are given in
Table~\ref{tab:models}. Figure~\ref{fig:acmodels} and
Fig.~\ref{fig:gmodels} display the density and ionization structure as
well as the velocity field of the selected models. Models A and C are
the results of simulations with the same moderate stellar wind and the
same radiation source. But in the simulation leading to model A the
diffuse {\sc uv} radiation field originating from scattering on dust grains
was completely neglected. For that reason we got a higher
photoevaporation rate $\dot{M}_{\rm ph}$ for model C. In
Fig.~\ref{fig:acmodels} this is recognizable by the greater overall
density in the ionized regions and by the higher velocity in the
``shadow'' regions of the disk in the case of model C. In order to
investigate the variation of spectral characteristics with the stellar
wind velocity we chose the models with the greatest wind velocities
G2, G3 and G4. Figure~\ref{fig:gmodels} shows the increasing opening
angle of the cone of freely expanding wind with increasing wind
velocity.
 
\subsection{Strategy of solution}

\begin{figure*} 
%
\begin{center}
\epsfig{file=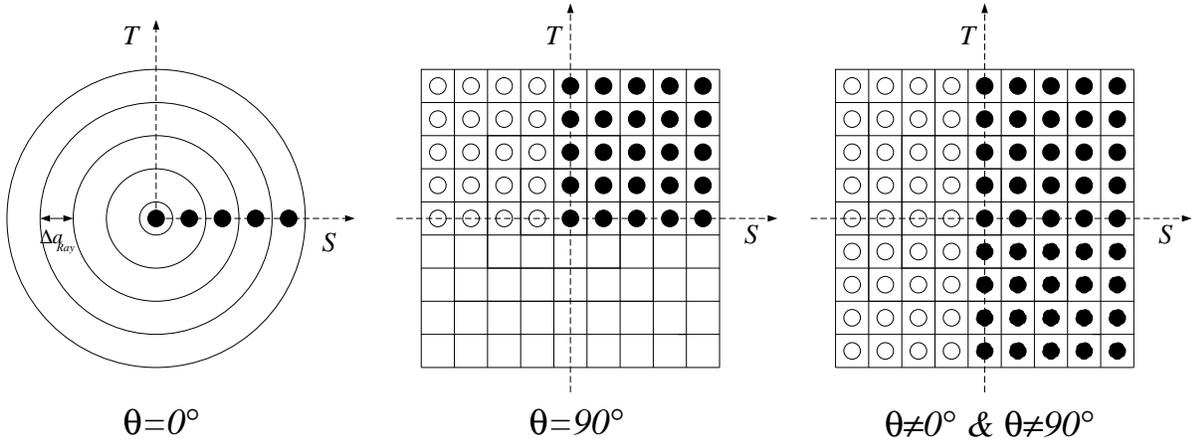,height=6.0cm}
\end{center}
\caption[]{Choice of lines of sight (LOS) and their associated areas for
 different viewing angles $\Theta$. Filled dots indicate the LOS
 used for the continuum calculations. Empty dots refer to the additional
 Lines of Sight necessary for the line profile calculations.}
  \label{fig:parcels}
\end{figure*}

We use the model data to calculate the ionization structure and
the level population. From the level populations we
determine the emissivities of each line transition and the
continuum emission at each point within the volume of the hydrodynamic mo\-del.
For each viewing angle $\Theta$ considered, we solve the time independent
equation of radiation transfer in a non-relativistic moving medium
along a grid of lines of sight (LOS) through the domain, neglecting
the effects of scattering:

\begin{equation} \label{eq:rte}
        \frac{{\rm d}I_\nu}{{\rm d}\tau_\nu}=-I_{\nu}+S_{\nu}  ,
\end{equation}
where the optical depth is defined as $\tau_\nu=\int \kappa_\nu {\rm d}s$.
Integrations were performed for a given set of frequencies, whereby the
effects of Doppler shifts for the line emissivities were taken into account.
The resulting intensities are used to determine SEDs,
intensity maps and line profiles. Spectra are obtained
from the spatial intensity distributions by integration, taking into
account that each LOS has an associated ``area''. Depending on $\Theta$
the symmetry of the configurations could be utilized to minimize the
computational effort (see Fig.~\ref{fig:parcels}). For the pole-on
view ($\Theta = 0^\circ$), for example, only a one dimensional LOS array
need be considered. For the edge-on view 
($\Theta = 90^\circ$) lines of sight either through a single quadrant
(continuum transfer) or through two quadrants (line transfer) are
necessary. The resolution of the central regions is enhanced by overlaying
a finer LOS grid in accordance with the multiple nested grid strategy
used in the hydrodynamic calculations. 

\begin{figure} 
\begin{center}
\epsfig{file=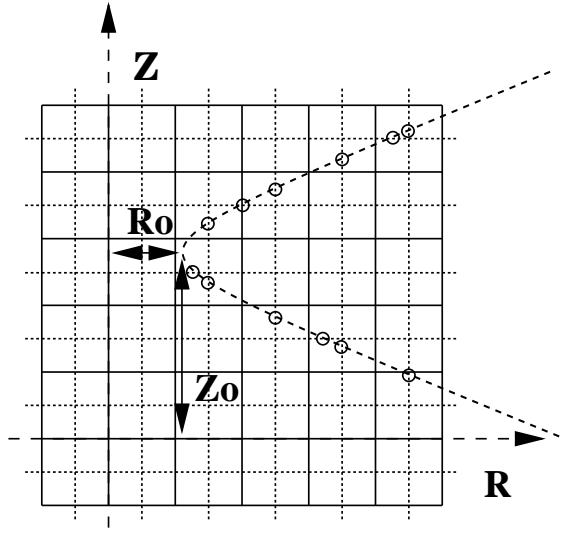,height=7.2cm}
\end{center}
\caption{Projection of a typical LOS trajectory (curved dashed line)
 onto the model data grid (solid lines). Temperature, density, degree
 of ionization and velocity are defined at cell centers. The small
 circles divide the LOS into subintervals; the source function $S_\nu$
 is evaluated at the location of the circles, chosen to lie on the
 intersections of the LOS with lines connecting the grid cell centers.}
  \label{fig:path}
\end{figure}

Each point in Fig.~\ref{fig:parcels} corresponds to an LOS trajecto\-ry
through the model. Mapping such a trajectory onto the (R,Z) model grid
yields hyperbolic curves as displayed in Fig.~\ref{fig:path}. 
Beginning with a starting intensity ($I_\nu(-\infty)$), the
solution of Eq.~(\ref{eq:rte}) is obtained by subdividing the
LOS into finite intervals and analytically integrating over each
interval assuming a sub-grid model (see below).

\subsection{Continuum radiation transfer}

If no discontinuities are present within the subinterval under consideration,
we assume $S_\nu$ varies linearly with $\tau$, i.e.\
\begin{equation}
  S_\nu(\tau)=S_\nu^i+(S_\nu^{i+1}-S_\nu^i) \frac{\tau}{\Delta \tau}  ,
\end{equation}
where $i$ and $i+1$ denote the starting and end points of the
interval, respectively, and $\Delta \tau$ is a mean optical depth over
the interval:
\begin{equation}
        \Delta \tau = \frac{\kappa^i + \kappa^{i+1}}{2} \Delta s  .
\end{equation}
With this formulation the solution of Eq.~(\ref{eq:rte}) over the entire
interval is given by (see Yorke~\cite{yorke1}):

\begin{eqnarray} \label{eq:step}
  I_{i+1} = I_{i} \exp ( - \Delta \tau) & + & S_{i} \left[ \frac{ 1-\exp
 (-\Delta \tau)}{\Delta \tau} -\exp (-\Delta \tau) \right] 
\nonumber \\
& + & S_{i+1} \left[ 1-\frac{1-{\rm exp}(-\Delta \tau)}{\Delta \tau} \right]
   .
\end{eqnarray}

For the cases considered here we choose $I_0 = 0$ as the starting LOS
intensity. For ``proplyd''-type models (considered in a subsequent paper
of this series) a non-negligible background intensity should be specified.

\subsection{Radiation transfer in emission lines}

For the transitions considered here the radiation field can be considered
``diffuse'' $I_\nu \ll B_\nu$ and the contribution of spontaneous
emission dominates over line absorption and stimulated emission processes.
After separating the source function $S_\nu = S_{\rm C} + S_{\rm L}$ and
the intensity $I_\nu = I_{\rm C} + I_{\rm L}$ of
Eq.~(\ref{eq:rte}) into the contributions of
the continuum and the line, we obtain
\begin{equation} \label{eq:linint}
   I_{\rm L} = \frac{\tilde{S}_{\rm L} \exp (-\Delta \tau)}{ \sqrt{\pi}
   \Delta \nu_{\rm th}} \int_0^{\Delta \tau}\exp\left[\tau-\left(
\frac{\nu - \tilde{\nu}(\tau)}{\Delta \nu_{\rm th}} \right)^2 \right]
   {\rm d} \tau ,
\end{equation}
where ${\rm d}\tau = \kappa_{\rm C} \, {\rm d}s$
and $S_{\rm L}=\epsilon_{\rm L}/ \kappa_{\rm C}$.
Here $\tilde{\nu}(\tau)$ is the Doppler-shifted frequency of the
transition, $\Delta \nu_{\rm th}$ the Doppler width and $\tilde{S}_{\rm L}$
the net source function integrated over the line. Assuming that
$\tilde{\nu}$ is linear in $\tau$ over the whole interval yields the
analytical solution to Eq.~(\ref{eq:linint}):
\begin{equation} \label{eq:linsol}
  I_{\rm L} = \frac{\tilde{S}_{\rm L} \Delta \tau}{2 \Delta \tilde{\nu}} \exp
\left( - \frac{(
\tilde{\nu}_2 - \nu) \Delta \tau}{\Delta \tilde{\nu}} \right)
\cdot \left[ \mbox{erf} (Y_2) - \mbox{erf} (Y_1) \right] ,
\end{equation}
where ${\rm erf}(y)=2/\sqrt{\pi} \int_0^z \exp(-t^2) dt$ is
the error function and $Y_i=(\tilde{\nu}_i-\nu)/\Delta \nu_{\rm th}$
a dimensionless frequency shift.

The net source function $\tilde{S}_{\rm L}$ is calculated according to the
algorithm suggested by Yorke (\cite{yorke1}):
\begin{eqnarray}
  \tilde{S}_{\rm L} & = & \frac{ \mbox{erf}(Y_{\rm M}) - \mbox{erf}(Y_{\rm
1})}{\mbox{erf} (Y_2) - \mbox{erf} (Y_1)} \tilde{S}_1 
+ \frac{ \mbox{erf} (Y_2) - \mbox{erf}(Y_{\rm M})}{\mbox{erf}(Y_2) -
\mbox{erf}(Y_1)} \tilde{S}_2,
\end{eqnarray}
with $Y_{\rm M} = (Y_1 + Y_2) / 2$. The line source functions $\tilde{S}_1$
and $\tilde{S}_2$ are calculated from the total line emission coefficient
$\tilde{\epsilon}_{\rm L}$ as defined in Eq.~(\ref{eq:lineemis}) at the
boundaries of the evaluated interval and from the continuum absorption
coefficient: $\tilde{S}_{\rm i}= \tilde{\epsilon}_{\rm L,i}/{\kappa_{\rm C}}$.

If $\Delta \tilde{\nu} \ll \Delta \nu_{\rm th}$, i.e.\ there is negligible
Doppler shift within the subinterval, the solution of
Eq.~(\ref{eq:linint}) with $\tilde{\nu}_1=\tilde{\nu}_2=\tilde{\nu}$ is used:
\begin{equation} \label{eq:narrowl}
  I_{\rm L} = \frac{S_{\rm L}}{\sqrt{\pi}\Delta \nu_{\rm th}} \exp \left[ - \left(
  \frac{\nu-\tilde{\nu}}{\Delta \nu_{\rm th}} \right)^2 \right] \left(1-\exp(-\Delta \tau)
  \right) .
\end{equation}

\subsection{Treatment of ionization fronts}

The numerical models considered contain unresolved ionization fronts
due to the coarseness of the hydrodynamic grid. At these positions jumps
occur in the physical parameters and the solutions given by
Eq.~(\ref{eq:step}) and Eq.~(\ref{eq:linsol}/\ref{eq:narrowl})
are poor approximations.
The exact location of the fronts within a grid cell are unknown; we
assume they lie at the center of the corresponding interval. Our criterion
for the presence of an ionization front is a change in the degree
of ionization $\Delta x > 0.1$ between two evaluation points. 

For the continuum calculations Eq.~(\ref{eq:step}) is applied to each
half interval with $\Delta \tau = \kappa_i \Delta x /2$. For the first
half $S= S_i$ kept constant and for the second half $S=S_{i+1}$ is
held constant.
For the line calculations Eq.~(\ref{eq:narrowl}) is used
with $\tilde{\nu}=\nu_1$ ($\nu_2$) and $S_{\rm L}=S_1$ ($S_2$) 
for the first (second) half interval.

\subsection{Treatment of the central radiation source}

The central source is modeled by a black body radiator of temperature $T_*$
and radius $R_*$. The integration along the line of sight through the center is
started at the position of the source with the initial intensity
\begin{equation}
  I_0 = B_\nu(T_*) \frac{R_*^2 \pi}{A}  ,
\end{equation}
with $A$ is the area associated with the central LOS.

\section{Results} \label{results}

With the code described above we determined SEDs, continuum isophotal
maps and line profiles for the forbidden lines [N{\sc ii}] 6584,
[O{\sc ii}]~3726 and [O{\sc iii}] 5007 as well as the H$\alpha$-line
for the models introduced in Sect.~\ref{intmod}. Their relevant
physical parameters are listed in Table~\ref{tab:models}.

\subsection{Continuum emission}

\subsubsection{Spectral Energy Distributions}

\begin{figure}
\begin{center}
\epsfig{file=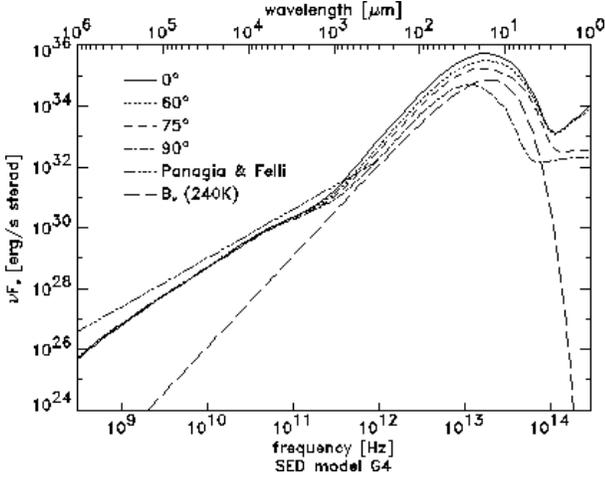,height=6.25cm}
\end{center}
\caption[]{Spectral Energy Distribution for model G2 and different $\Theta$.}
\label{fig:modg2cont}
\end{figure}

Figure~\ref{fig:modg2cont} shows the SEDs of model G2 presented in Paper~II for different viewing angles $\Theta$.
The spectra can be divided into three regimes dominated by different
physical processes:
\begin{enumerate}
\item In the frequency range from $10^8$ to $10^{11}$ Hz the SED is
dominated by the thermal free-free radiation in the ionized region
around the dust torus.
\item The {\sc ir}-excess from $10^{11}$ to $10^{14}$ Hz is due to the
optically thick dusty torus itself, which has a mean surface temperature of
about 250 K.
\item Beyond $10^{14}$ Hz the SED depends strongly on the viewing
angle: if the star is obscured by the dusty torus then the free-free
radiation of the H{\sc ii}-region again dominates the spectrum,
otherwise the stellar atmosphere shows up. Due to the uncertainties in
the stellar spectra and the neglect of scattering by dust in
Eq.~(\ref{eq:rte}), which becomes more and more important with
increasing frequency, a discussion of the SED beyond $10^{14}$~Hz and
comparison with observations are not useful.
\end{enumerate}

\begin{table*}[tb]
 \caption{Scattering coefficient
  $\kappa^{\rm scat}_{\rm dust} \rho^{-1}$ as well as parameters for
  the stellar wind (mass loss rate $\dot{M}_{\rm wind}$ and
  velocity $v_{\rm wind}$) and the ionizing source (stellar photon
  rate $S_{\rm star}$ and temperature $T_{\rm eff}$) used in the
  calculations. The evaporation time scale $t_{\rm evap}$ is
  calculated from
  $t_{\rm evap}=M_{\rm disk}/\dot{M}_{\rm ph}$ with
  $M_{\rm disk} = 1.67\,M_\odot$.}
  \begin{center}
 \begin{tabular}{cccccccc}
 \hline\noalign{\smallskip}
   model & $\kappa^{\rm scat}_{\rm dust}/\rho$ & $\dot{M}_{\rm wind}$
  & $v_{\rm wind}$ & $\log_{10} S_{\rm star}$ & $T_{\rm eff}$
  & $\dot{M}_{\rm ph}$ & $t_{\rm evap}$ \\

   &  $\mbox{cm}^2\mbox{g}^{-1}$ & $10^{-8}M_\odot\mbox{yr}^{-1}$ & $\mbox{km\,s}^{-1}$ &
   $\mbox{s}^{-1}$ & $\mbox{K}$ & $10^{-6}M_\odot\mbox{yr}^{-1}$ &
   $10^6\mbox{yr}$ \\ 
 \noalign{\smallskip}
 \hline\noalign{\smallskip}    
  A & 0 & 2 & 50 & 46.89 & 30\,000 & 0.565 & 2.96 \\
  C & 200 & 2 & 50 & 46.89 & 30\,000 & 1.35 & 1.24 \\
 \hline\noalign{\smallskip}
  G2 & 200 & 2 & 400 & 46.89 & 30\,000 & 1.65 & 1.01 \\
  G3 & 200 & 2 & 600 & 46.89 & 30\,000 & 1.67 & 1.00 \\
  G4 & 200 & 2 & 1000 & 46.89 & 30\,000 & 1.64 & 1.02 \\
 \noalign{\smallskip}
 \hline
 \end{tabular}
 \label{tab:models}
 \end{center}
\end{table*}

According to the analysis of Panagia \&\ Felli~(\cite{panagia}), who
calculated analytically the
free-free emission of an isothermal, spherical, ionized wind,
the flux density should obey a $\nu^{0.6}$-law:
\begin{eqnarray}
 F_{\nu}=6.46 \cdot 10^{-3}\; \mbox{Jy} \cdot \left[ \frac{\dot{M}}{10^{-5} \mbox{~M}_\odot / \mbox{yr}} \right] ^{4/3} \cdot
  \left[ \frac{\nu}{10\mbox{~GHz}} \right] ^{0.6} \cdot \nonumber \\ 
  \cdot \left[ \frac{T}{10^4\mbox{~K}} \right] ^{0.1} 
  \cdot \left[ \frac{d}{\mbox{kpc}} \right] ^{-2} \cdot 
  \left[ \frac{v_{\rm wind}}{10^3\mbox{~km\,s$^{-1}$}} \right] ^{-4/3}.
\label{eq:panagia}
\end{eqnarray}
Schmid-Burgk~(\cite{schmid}) showed
that this holds, modified by a geo\-metry dependent factor of order unity, even for non-symmetri\-cal point-source winds as long as $\rho$ drops as $R^{-2}$. Additionally he postulated that the flux
density should hardly be dependent on the angle at which the object is viewed.
In Fig.~\ref{fig:modg2cont} we include for comparison the flux distribution
given by Eq.~\ref{eq:panagia} for the photoevaporation rate
$\dot{M}_{\rm ph}=1.65\cdot
10^{-6}\;\mbox{M}_\odot\mbox{yr}^{-1}$ (see Paper II),
$T=10\,000$~K and $v_{\rm wind}=20\;\mbox{km~s}^{-1}$ derived from the line profiles in Sect.~\ref{sect:lines}.
The slope of the SED in regime 1 is slightly steeper in our results,
because our volume of integration is finite;
Panagia \&\ Felli~(\cite{panagia}) derived their analytical results by assuming an infinite
integration volume. The flux is almost independent of $\Theta$, which
is in good agreement with Schmid-Burgk~(\cite{schmid}). The deviations between $10^{10}$ and $10^{11}$ Hz
are due to the break in the $R^{-2}$-law caused by the neutral dust torus.

Figure~\ref{fig:modg2cont} also includes the SED of a blackbody at $T=240$~K. The
flux $F_{\nu} \propto \nu^{2.2}$ in the far {\sc ir} between $5\cdot10^{11}$\,Hz and
$3\cdot10^{12}$\,Hz is slightly steeper than the comparison
blackbody spectrum, because the dust torus is not quite optically
thick. With increasing $\Theta$ the maximum shifts towards lower
frequencies, because the warm dust on the inside of the torus is being
concealed by the torus itself.

We obtain qualitatively the same results for a number of models
presented in Paper II.

\begin{figure*}
\begin{center}
\epsfig{file=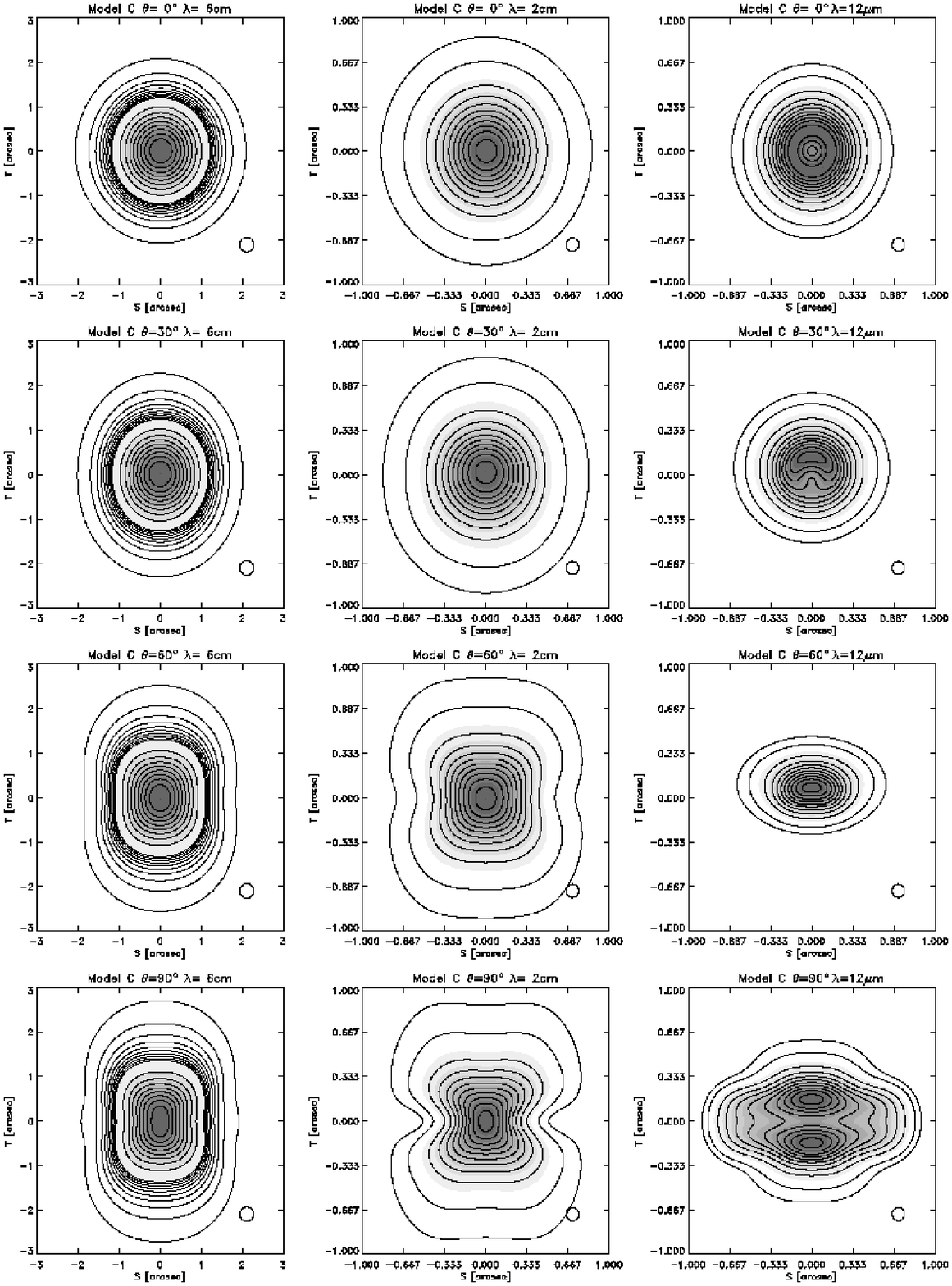,height=21cm}
\end{center}
\caption[]{Isophotal maps of model C for different viewing angles and
wavelengths as indicated. Assumed distance 300 pc. The circle in the
lower right corner marks the FWHM of the point spread function. Values
for contours see text.}
\label{fig:contourc}
\end{figure*}

\begin{figure*}
\begin{center}
\epsfig{file=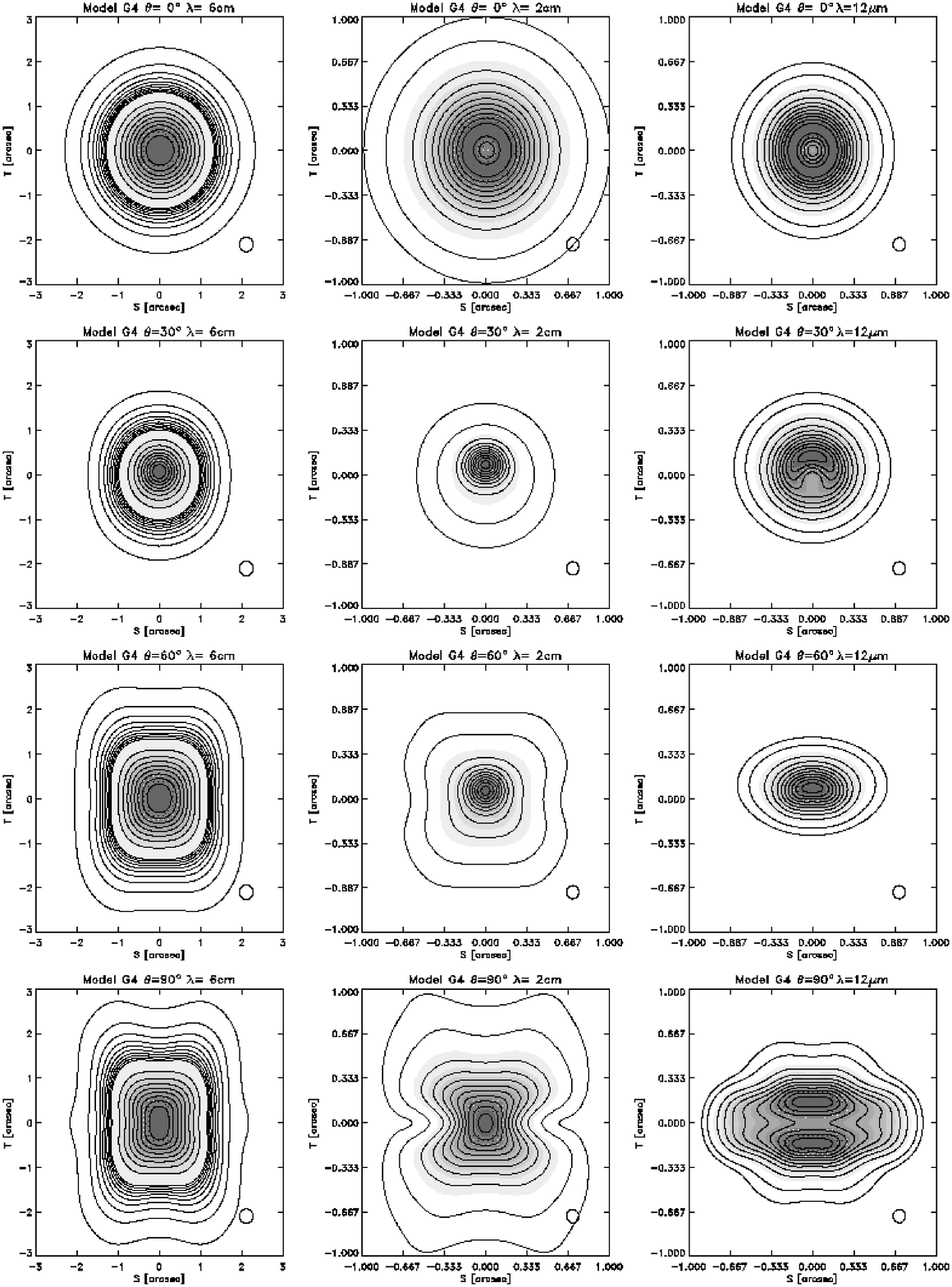,height=21cm}
\end{center}
\caption[]{Isophotal maps of model G4 for different viewing angles and
wavelengths as indicated. Assumed distance 300 pc. The circle in the
lower right corner marks the FWHM of the point spread function. Values
for contours see text.} 
\label{fig:contourg4}
\end{figure*}

\subsubsection{Isophotal maps}

We also calculated isophotal maps over the projected $(S,T)$ grid
of the sky for models C and G4
(Figs.~\ref{fig:contourc},\ref{fig:contourg4}). The maps were
convolved with a Gaussian point spread function (FWHM 0\mysec3 for
$\lambda=6$\,cm, 0\mysec1 for $\lambda=2$\,cm and $12\,\mu$m) 
in order to compare
the numerical models with observations of limited resolution. The
values (in percent) of the contour lines relative to the maximum flux 
per beam are 1, 2, 3, 4, 5, 6, 7, 8, 9, 10, 20, 30, 40, 50, 60, 70, 80,
90 for $\lambda=6\,$cm, and 2, 5, 15, 25, 35, 45, 55, 65, 75, 85, 95
for $\lambda=2\,$cm and $12\mu$m. 

The difference in these two
models is revealed most strikingly in the maps for $\lambda=6\,$cm and
$2\,$cm. The mass outflow of model C is more evenly distributed over
its whole opening angle, whereas in model G4 most of the mass is
transported outward in a cone between $\Theta=30^\circ$ and
$70^\circ$. This leads to the X-shape of the corresponding radio maps
for viewing angles $\Theta \ge 60^\circ$. The high density in the
region between star and disk for this model results in an optically
thick torus in this region at
$\lambda=2\,$cm. Thus the contours for $\Theta=30^\circ$ and
$60^\circ$ are not symmetric to the equatorial plane at $T=0^{''}$.  

In the maps corresponding to $\Theta = 30^\circ, \lambda = 12 \mu$m
there appears a peculiar horseshoe-like feature. It is generated by
the hottest region of the dust torus, which is the innermost boundary
with the smallest distance to the star. It can be seen as a ring in
the maps for $\Theta = 0^\circ$. For $\Theta = 30^\circ$ the part of the
ring next to the observer is obscured by the dust torus, whereas the
other parts are still visible. The beam width chosen by us allows the
resolution of the ring and thus leads to the horseshoe-like
feature. For $\Theta = 60^\circ$ only the most distant part of this hot
ring is visible, leading to a maximum in the flux with a smaller spatial
extent than for $\Theta = 30^\circ$.

\begin{figure*}
\begin{center}
\epsfig{file=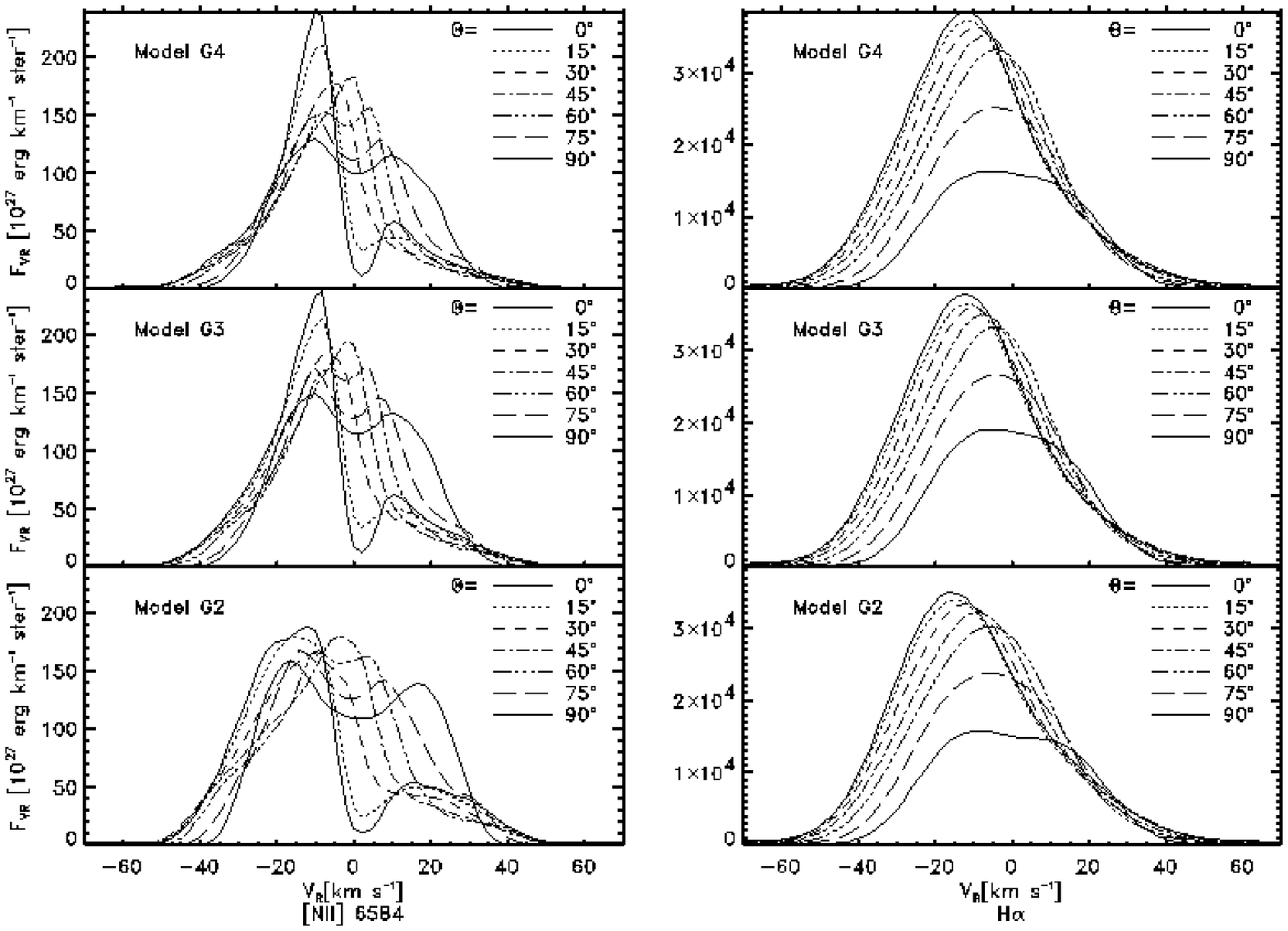,height=17cm,angle=270}
\end{center}
\caption[]{Lines [N{\sc ii}] 6584 and H$\alpha$ for models G2, G3 and
G4, and different ``viewing angles'' $\Theta$. The models differ only
in wind velocity $v_{\rm wind}$.}
\label{fig:lines1}
\end{figure*}

\begin{figure*}
\begin{center}
\epsfig{file=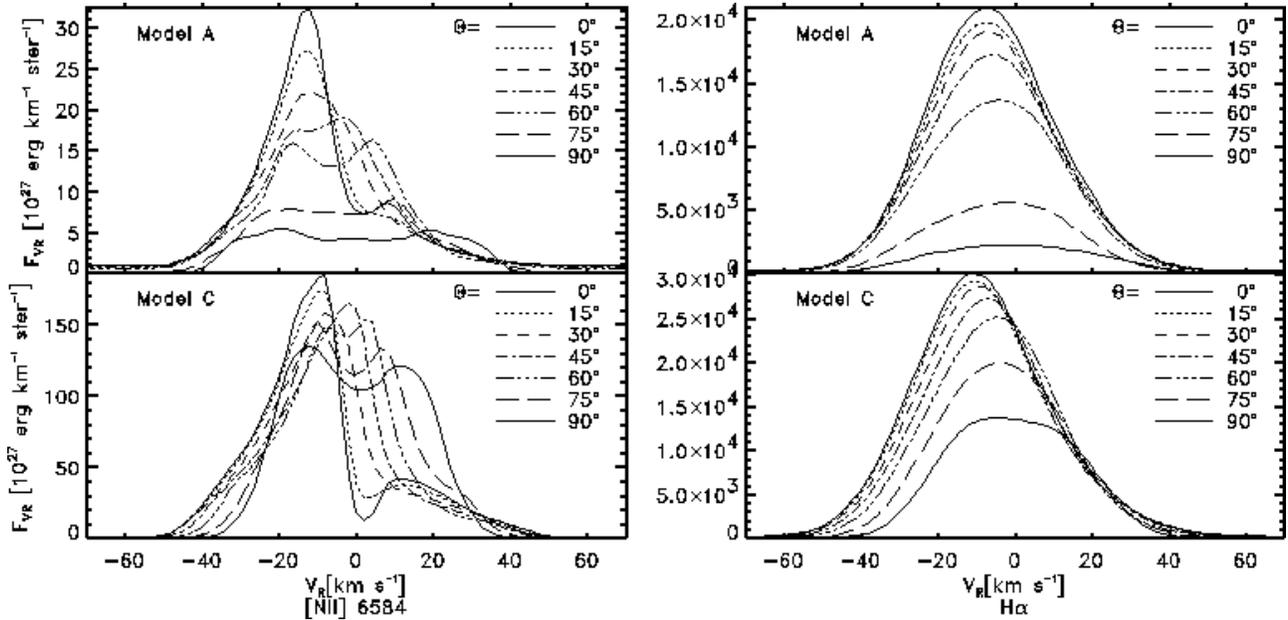,height=17cm,angle=270}
\end{center}
\caption[]{Lines [N{\sc ii}] 6584 and H$\alpha$ for models A and
C, and different ``viewing angles'' $\Theta$. In model A scattering of
{\sc uv} photons by dust was neglected, in model C included. Note the
different scales on the abscissae.}
\label{fig:lines3}
\end{figure*}

\begin{figure*}
\begin{center}
\epsfig{file=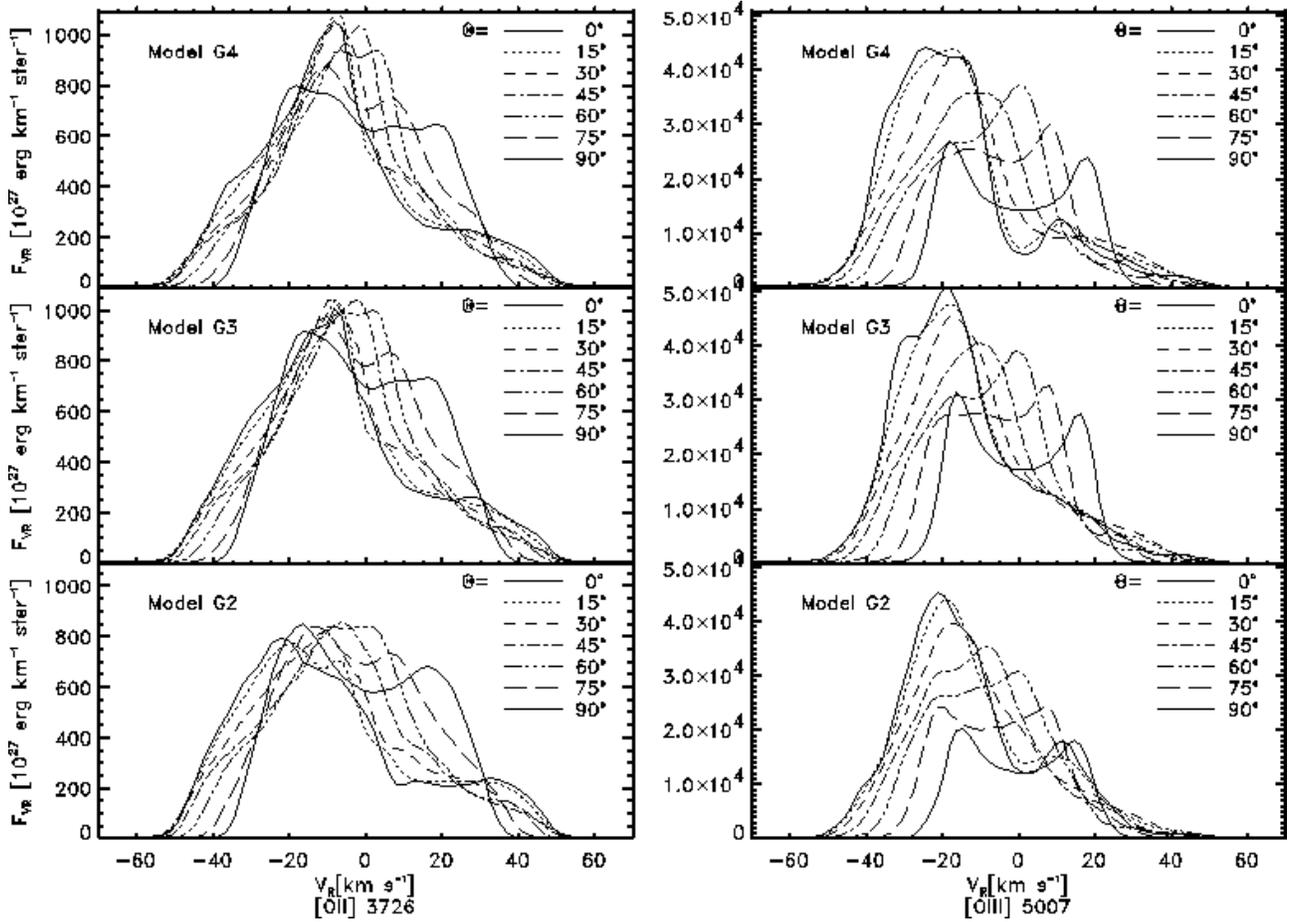,height=17cm,angle=270}
\end{center}
\caption[]{Lines [O{\sc ii}] 3726 and [O{\sc iii}] 5007 for models G2,
G3 and G4, and different ``viewing angles'' $\Theta$. The models differ only
in wind velocity $v_{\rm wind}$.}
\label{fig:lines2}
\end{figure*}

\begin{figure*}
\begin{center}
\epsfig{file=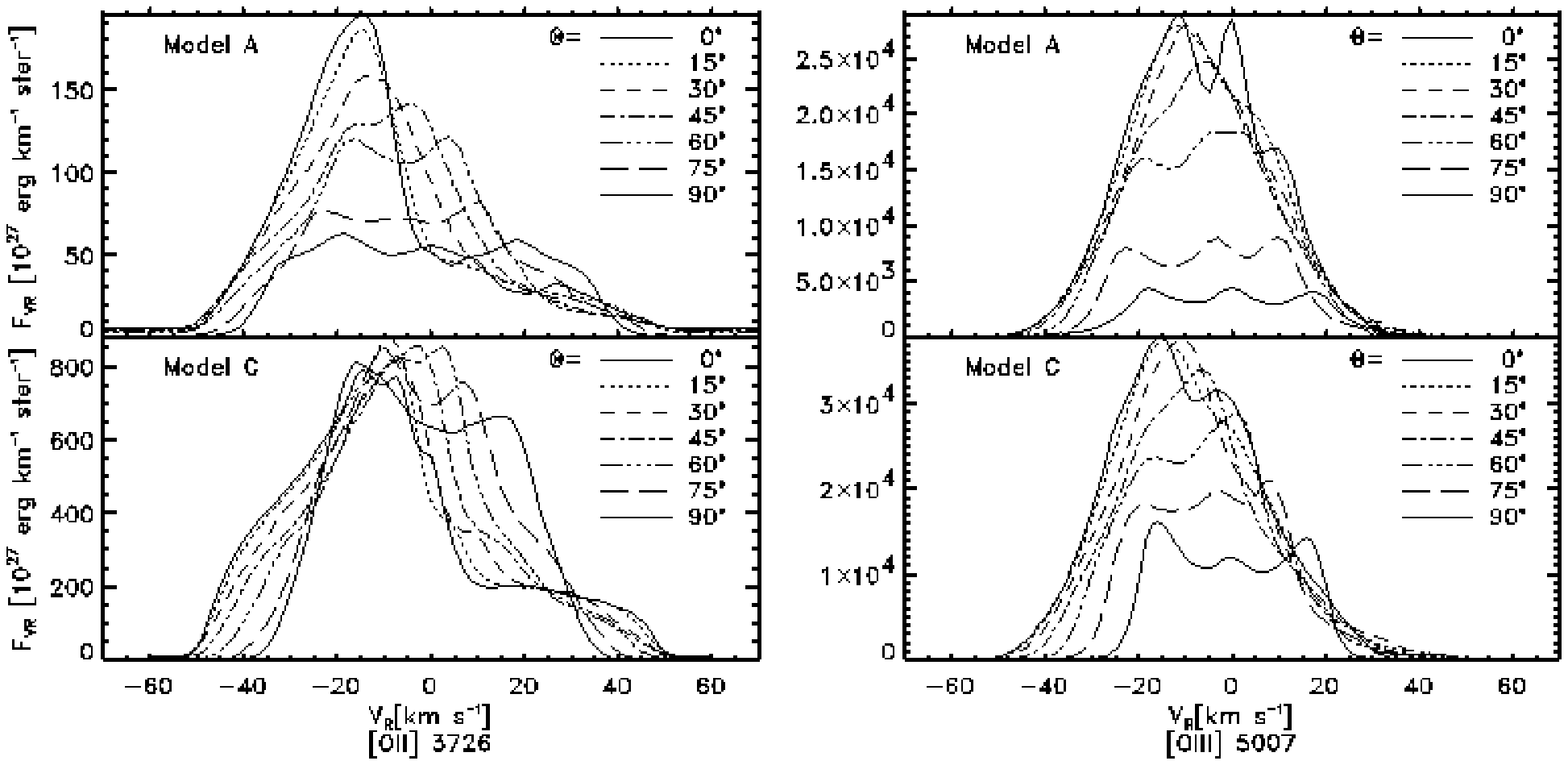,height=17cm,angle=270}
\end{center}
\caption[]{Lines [O{\sc ii}] 3726 and [O{\sc iii}] 5007 for models A
and C, and different ``viewing angles'' $\Theta$. In model A scattering of
{\sc uv} photons by dust was neglected, in model C included. Note the
different scales on the abscissae.}
\label{fig:lines4}
\end{figure*}

\subsection{Line profiles} \label{sect:lines}

Line profiles can provide useful information on the velocity
structure in H{\sc ii} regions. Because the thermal doppler broadening
decreases with increasing atomic weight as $A^{-1/2}$, ions such as
N{\sc ii} and O{\sc iii} are better suited for velocity diagnostics
than H$_\alpha$. This can be seen in the line
profiles we obtained (Figs.~\ref{fig:lines1}-\ref{fig:lines4}). They
show the flux integrated over the whole area of the object including
the disk, the evaporated flow and the cone of the stellar wind. In all cases the line
broadening of several ten km\,s$^{-1}$ is dominated by the velocity distribution of the
escaping gas. For comparison, the rotational velocity of the dust
torus $v_{\rm rot} \simeq 13$~km\,s$^{-1}$ for the inner parts, and the thermal Doppler broadening from
Eq.~(\ref{eq:doppler}) at $T = 10\,000$~K is $v_{\rm th} \simeq 9$~km\,s$^{-1}$
for H$\alpha$ and $\simeq 2$~km\,s$^{-1}$ for O{\sc iii}.

Figures~\ref{fig:lines3} and \ref{fig:lines4} show the line profiles
for the models A (no scattering of H-ionizing photons)
and C (calculated assuming non-negligible {\sc uv} scattering during
the hydrodynamical evolution) 
from Paper~II. {\sc uv} scattering leads to stronger illumination of the
neutral torus by ionizing radiation and thus to a 
higher photoevaporation rate in model C (by a factor of $\sim 2.3$)
compared to model A. Due to
the higher density in the regions filled with photoevaporated gas, the
lines for model C are generally more intense. In the case of the line
[O{\sc ii}] 3726 one notices that the difference between the fluxes
for different angles is the smallest of all transitions. Not only is the
density of the outflowing ionized gas higher for case~C, but the charge
exchange reactions discussed in Sect.~\ref{sect:chargeex} lead to
the establishment of a non-negligible O{\sc ii}-abundance even in the
``shadow'' regions not accessible to direct stellar illumination.
These regions dominate the line spectra for all angles.

Figures~\ref{fig:lines1} and \ref{fig:lines2} show the calculated line
profiles for the models G2, G3 and G4, which only differ by the
assumed stellar wind velocity $v_{\rm wind}$, increasing from
400~km\,s$^{-1}$ (G2) to 1\,000~km\,s$^{-1}$ (G4). Comparing the
results we find two features:
\begin{enumerate}
\item The intensity and overall structure of the line profiles considered
are almost independent of the stellar wind velocity $v_{\rm wind}$.
\item No high-velocity component appears in the lines due to the
stellar wind.
\end{enumerate}

Both features can be explained by the fact that the density of material
contained in the stellar wind is much lower than the density in the 
photoionized outflow. Remembering that $\rho = \dot M / 4\pi r^2 v$ in
a steady-state outflow and that the line emissivity
$\epsilon_{\rm L} \propto \rho^2$, we can understand why the low
expansion velocities of about $20$~km\,s$^{-1}$ (i.e. material close
to the torus) dominate the spectra.
The overall evaporation rates as well as the
expansion velocities are almost equal for all three models, leading
to very similar line profiles. It would be necessary to consider 
transitions which predominate in hot winds in order to detect this
high velocity component and to find significant differences between
these models.

In spite of our assumption of optically thin line transfer, the
profiles for $\Theta=90^\circ$ are not symmetric. This is due to the
dust extinction within the H{\sc ii}-region. The receding material is
on average further away from the observer than the approaching. The
longer light paths result in stronger extinction of the redshifted 
components.

We are aware of the fact that the neglection of scattering by dust in
Eq.~(\ref{eq:rte}) may lead to serious errors in the calculated
line profiles. We expect non-negligible contributions especially in
the red-shifted parts of the lines due to light scattered by the
dense, dusty surface of the neutral torus. This light was originally
emitted by gas receding from the torus. The resulting redshift ``seen''
by the torus remains unchanged during the scattering process and will thus
lead to enhancement of the red-shifted wings of the lines. 

Nevertheless, we expect our qualitative results still to hold since
the arguments mentioned above referring to the geometry of the
underlying models are still applicable.

\section{Comparison with observations} \label{comparison}

Although the cases presented here describe the situation of an 
intermediate mass ionizing source (8.4 M$_\odot$) with a circumstellar 
disk, many of the basic spectral features can be generalized. Thus,
it is interesting to compare and contrast our results with 
observed UCH{\sc ii}s, even though many of the central sources are 
presumably much more massive.

The collections of photometry data in the catalogues of Wood \&\
Churchwell~(\cite{wood}) and Kurtz et al.~(\cite{kurtz})
show that the SEDs of almost all
UCH{\sc ii}-regions possess roughly the same structure as the ones of
the models discussed here: a flat spectrum in the radio and mm regime
following a $\nu ^{0.6}$-law due to free-free emission and an {\sc ir} excess
originating from heated dust
exceeding the free-free emission by $\sim 3-4$ orders of magnitude. A
closer inspection shows that the dust temperatures are by an order of
magnitude lower in most of the observed sources when compared to
our models. This may be an indication that the disks are being 
photoionized by a close companion in a multiple system rather than
the central source. Alternatively, the emitting dust could be
distributed in a shell swept up by the expanding H{\sc ii}-region and
thus further away from the exciting star than for the cases discussed
here. The large
beam width of IRAS and the tendency of massive stars to form in
clusters also make it likely that the {\sc ir} fluxes belong to
dust emission caused by more than one heating star. 
Objects of this type would
appear as ``unresolved'' in the maps presented by
Wood \&\ Churchwell~(\cite{wood}) for distances larger than
$\sim$~300~pc. Due to the cooler dust, the ``unresolved'' objects cannot be
explained by the models of circumstellar disks around {\sc uv} luminous
sources specifically discussed in this paper. Certainly the cometary
shaped UCH{\sc ii}s can be explained by a disk being evaporated by the
ionizing radiation of an external star and interaction with its
stellar wind. Numerical models dealing with this scenario will be 
presented in the next paper of this series.
\begin{figure*}
 \begin{center}
   \epsfig{file=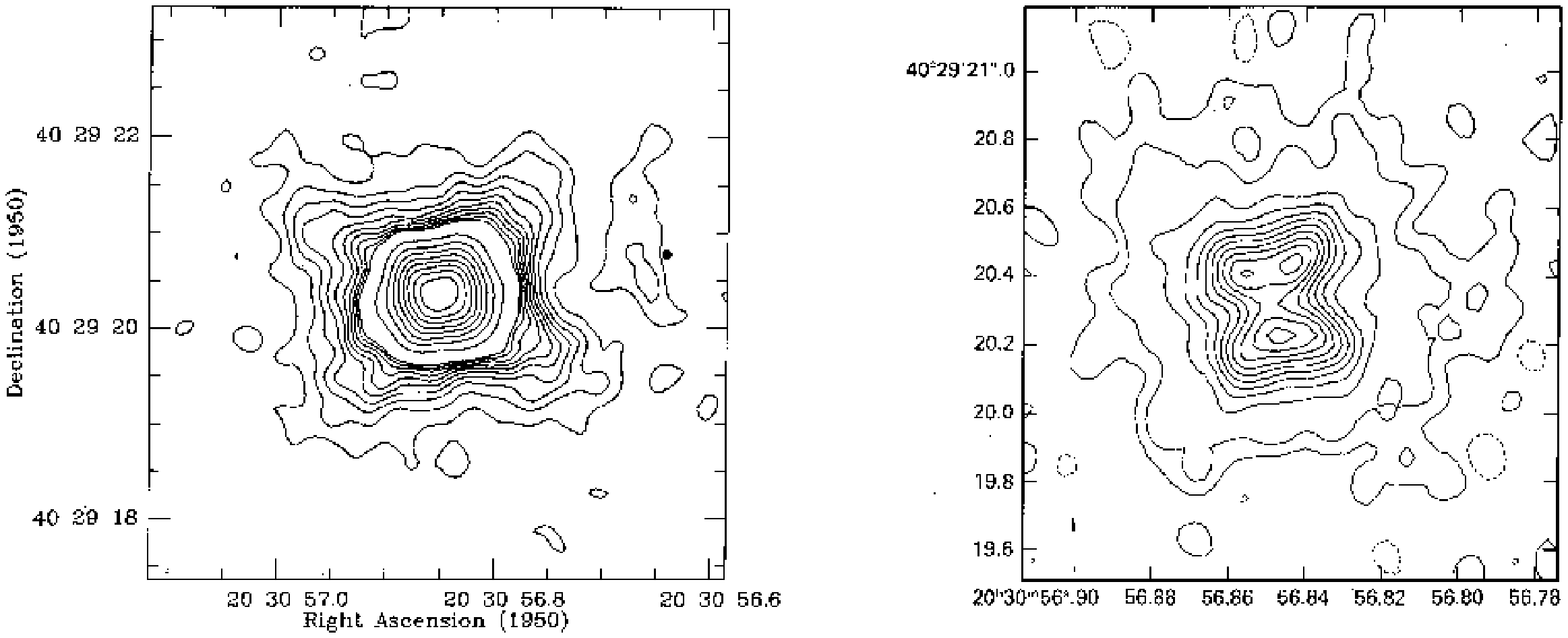,width=5.5cm,angle=270}
 \end{center}
 \caption{VLA-maps of \object{MWC~349}. Left: $\lambda=$~6~cm (Cohen et
 al. \cite{cohen}), FWHM~$=$~0\mysec3. Contour levels at 1, 2,..., 9,
 10, 20,..., 80, 90\% of maximum flux 16,59~mJy\,beam$^{-1}$. Right:
$\lambda=$~2~cm (White \&\ Becker~\cite{white}),
 FWHM~$=$~0\mysec1. Contour levels at $-2$, 2, 5, 15, 25,...95\% of
 maximum flux 156~mJy\,beam$^{-1}$.}
 \label{fig:mwcvla}
\end{figure*}



\subsection{\object{MWC~349}~A}

A commonly accepted candidate for an evaporating disk around a young
massive star
is \object{MWC~349}~A. Its radio continuum flux obeys the $\nu^{0.6}$-law
up to $\lambda=30$~cm (see collection of photometry results in Thum
\&\ Mart\'{\i}n-Pintado~\cite{thum}). Radio maps obtained by Cohen et al.~(\cite{cohen}) and
White \&\ Becker~(\cite{white}) show an extended ionized bipolar
outflow with a peculiar X-shape (Fig.~\ref{fig:mwcvla}). In the center
Leinert~(\cite{leinert}) finds an elongated,
dense clump with $T \sim 900\,$K, with optically thick {\sc ir} emission,
which makes \object{MWC~349}~A one of the brightest IRAS sources. The elongated
structure shows an almost Keplerian velocity profile along its
major axis, perpendicular to the outflow axis (Thum \&\ Mart\'{\i}n-Pintado~\cite{thum}). This leads to the assumption
of a neutral dust torus around the central star with a small outer radius
$< 100\,$AU. Kelly et al.~(\cite{kelly}) estimated the extinction
towards this early-type star to be $A_{\rm V} = 10.8$.

The SED of \object{MWC~349}~A shows all the features
which we find for our models. The extinction and the geometry of the
outflow, as well as the lack of a high-velocity component in the line
profiles (Hartmann et al. \cite{hartmann}, Hamann \&\ Simon~\cite{hamann}), could be explained by a model with fast stellar
wind presented in this paper, assuming a viewing angle $\Theta \sim
90^\circ$. On the other hand, the high dust temperatures in the torus,
$T_{\rm d} \sim 800$\,K,
and the extremely high mass loss rate in the outflow, $\dot{M} = (1.16
\pm 0.05) \times 10^{-5}\,$M$_\odot$\,yr$^{-1}$ (Cohen et al. \cite{cohen})
remain puzzling and need clarification by a numerical model after the
method described in this series but especially ``tailored'' to suit the
needs of \object{MWC~349}~A.

\section{Conclusions} \label{sect:conclusions}

In this paper we showed that the models of photoevaporating disks
around intermediate mass stars cannot explain
the large number of ``unresolved'' UCH{\sc ii}'s observed by Wood \&\
Churchwell~(\cite{wood}) and Kurtz et al.~(\cite{kurtz}), because the
inferred dust temperatures of these objects are in most cases an
order of magnitude lower than those obtained in the numerical models.
But the question remains whether the disks of
more massive stars than considered here could be responsible for the
high abundance of the ``unresolved'' UCH{\sc ii}'s.
Disks around close companions of massive stars should be treated in
greater detail. 
If we assume that circumstellar disks are the
rule in the process of star formation, the simplicity and
straightforwardness of the model make it favorable compared to
alternative suggestions. The extremely high radiation pressure in the
vicinity of massive stars could lead to a larger distance between star
and disk and thus to smaller dust temperatures. 

Another important result of this work is that the profiles of
forbidden lines in the optical for the models G2, G3 and G4 with wind
velocities of $400-1000$ km\,s$^{-1}$ are almost independent of
$v_{\rm wind}$. This is due to the fact that the mass
loss rate and velocity of the evaporated disk material is not
affected by its interaction with the stellar wind, but by the rate
of ionizing photons and the peculiar shock structure which is
very similar in the numerical models (see Fig.~\ref{fig:gmodels}). 
Since the total
mass loss rate is dominated by the evaporated component with low
velocity, and the emission is proportional to $\rho^2$, the intensity
in the lines does not depend on details of the stellar wind
 and their profiles show the same low-velocity components.

Treatment of optically thick line emission and scattering effects is not possible with the
method presented above. In order to compare non-LTE-effects like
masing lines, which are observed in various objects related with the
formation of massive stars, one has to refer to different methods,
e.g. the Monte-Carlo method presented by Juvela~(\cite{juvela}). 
This would immensely help us in our
understanding of the process of formation and evolution of massive stars.

\begin{acknowledgements}
This research has been supported by the Deutsche
Forschungsgemeinschaft (DFG) under grants number Yo
5/19-1 and Yo 5/19-2. Calculations were performed at the HLRZ in J\"ulich and the
LRZ in Munich. We'd also like to thank D.J. Hollenbach for useful discussions.
\end{acknowledgements}

\end{document}